\documentclass[preprint,onecolumn,superscriptaddress,tightenlines,nofootinbib,12pt]{article}
\pdfoutput=1 
\usepackage{graphicx}
\usepackage{epsfig}
\usepackage{amsmath,latexsym,amssymb,hyperref,enumerate}
\usepackage{amsfonts}
\usepackage{url}
\usepackage{color}
\usepackage{cancel}
\usepackage{cite}
\hypersetup{colorlinks,citecolor= nicegreen,linkcolor= nicered}
\definecolor{nicered}{rgb}{0.7,0.1,0.1}
\definecolor{nicegreen}{rgb}{0.1,0.5,0.1}

\begin{document}
\renewcommand{\thefootnote}{\fnsymbol{footnote}}
\renewcommand{\baselinestretch}{1.2}
\normalsize
\vspace{-0.5cm}
\begin{flushright}
CALT-TH-2014-160\\
\vspace{1cm}
\end{flushright}

\vspace*{1.2cm}

\centerline{\bf\Large\textcolor[rgb]{0,0,0}{Yukawa Bound States of a Large Number of Fermions}}
%\vspace*{0.25cm}
%\centerline{\bf\Large and Application to Dark Matter}
\vspace*{1.5cm}

\centerline{{\bf Mark~B.~Wise}\ \ and \ \  {\bf Yue~Zhang}}

\vspace*{1.0cm}

\centerline{\em Walter Burke Institute for Theoretical Physics,}
\centerline{\em California Institute of Technology, Pasadena, CA 91125}

\vspace*{0.5cm}

\centerline{\tt wise@theory.caltech.edu,\ \ \ yuezhang@theory.caltech.edu}

\vspace{1.5cm}

\parbox{16.5cm}{
{\sc Abstract:} 
We consider the bound state problem for a field theory that contains a Dirac fermion $\chi$ that Yukawa couples to a (light) scalar field $\phi$.  We are interested in bound states with a large number $N$ of  $\chi$ particles. A Fermi gas model is used to numerically determine the dependence of the radius  $R$ of these bound states on $N$ and also the dependence of the binding energy on $N$. Since scalar interactions with relativistic $\chi$'s are suppressed two regimes emerge. For modest values of $N$ the state is composed of non-relativistic $\chi$ particles. In this regime as $N$ increases $R$ decreases. Eventually the core region becomes relativistic and the size of the state starts to increase as $N$ increases. As a result, for fixed Yukawa coupling and $\chi$ mass, there is a minimum sized state that occurs roughly at the value of $N$ where the core region first becomes relativistic. We also compute an elastic scattering form factor that can be relevant for direct detection if the dark matter is composed of such $\chi$ particles.
}

\newpage

\section{Introduction}

One of the mysteries of modern physics is the composition of the dark matter (DM). Various extensions of the standard model (SM) with dark matter candidates have been proposed and studied. A popular scenario has the Higgs boson mediating the connection between DM and the SM. This setup may be testable at  LHC experiments which explore Higgs physics. A very simple candidate for the DM is a Dirac fermion (which we denote by $\chi$) that is a singlet under the SM gauge group. If $\chi$ interacts with the SM sector through the Higgs boson, the lowest operator has dimension 5 which is not renormalizable.\footnote[4]{We neglect the renormalizable operator $\bar L H \chi$ which explicitly breaks the symmetry that stabilizes the DM. Including such interaction with a tiny coupling could lead to the decaying DM scenario and indirect detection signals.} If the cutoff scale is higher than that of the relevant physical processes, one can use this effective operator description. On the other hand, if the cutoff scale is lower than the DM mass, working with a ultraviolet (UV) completion is necessary. The simplest UV completion has an additional gauge singlet scalar (which we denote by $\phi$) that interacts with the DM through a Yukawa coupling $g_{\chi}$, and interacts with the Higgs boson via renormalizable couplings. For $m_\phi<m_\chi$, the Yukawa coupling allows the DM to efficiently annihilate in the early universe. This very simple dark matter sector and has been previously studied in the literature~\cite{Low:2011kp, Baek:2011aa, Djouadi:2011aa}. The scenario where DM self interacts with a light mediator has also drawn astrophysical interest~\cite{Spergel:1999mh, Rocha:2012jg, Buckley:2009in, Tulin:2013teo, Boddy:2014yra} because of its possible implications for structure formation on small scales.
 
One interesting aspect of the DM in this model is that for small enough scalar $\phi$ mass, the DM particles have a rich spectrum of $\chi$ particle bound states. Two-body bound states were studied in~\cite{rogers1970} and in the non-relativistic regime bound states with $N\gg1$ $\chi$ particles were studied in~\cite{Wise:2014jva}. For small values of the dark Yukawa coupling fine structure constant $\alpha_{\chi}= g_{\chi}^2/(4 \pi)$ and moderately large values of $N\lesssim \alpha_\chi^{-3/2}$ these bound states are non-relativistic. It was found that the size of these states decreases as $N$ is increased.

As $N$ increases the $\chi$ particles become more relativistic and eventually the methods used in  \cite{Wise:2014jva} are no longer applicable. The nature of the bound states for these larger $N$'s  is not known. The purpose of this paper is to fill in this gap by providing an understanding of the bound states with a large number of fermions where relativistic physics is important. While the main motivation for such states existing in nature is dark matter, the  results of this paper are not dependent on that physical interpretation.

One important difference between scalar interactions and vector interactions with fermions is that scalar Yukawa couplings give rise to interactions that are suppressed at large fermion energies. Because of this, we find that the character of the bound states changes as one enters the relativistic regime. They no longer get smaller as $N$ increases but rather start  growing in size. Eventually these bound states become so large that the screening of the Yukawa potential by the factor $e^{-m_\phi/r}$ cannot be ignored. 
%Larger stable bound states with more particles do not exist. 

In this paper we use both analytical and numerical methods to study the binding energy and the size of the Yukawa bound states as a function of $N$. We present our results in a number of plots and provide a detailed discussion of the methods used. Finally we provide a calculation of an elastic form factor that may be relevant for direct detection experiments if the $\chi$ particles are the dark matter.

\section{Fermions with a Yukawa Interaction}\label{method}

In this paper we study  bound states of $N$ Dirac fermions $\chi$ interacting through the exchange of a light scalar mediator $\phi$. We are interested in understanding the properties of such states at large $N$,\footnote[4]{We call such states nuggets.} in particular their binding energy and the dependence of the size of the states $R$ with the number of particles. Exchange of the light mediator results in an attractive potential between the $\chi$ particles. The range of the potential is set by the mass of the scalar $m_{\phi}$. 
%Since the Yukawa potential is screened at distances much larger than $1/m_{\phi}$, stable bound states with radius $R$ so large that $Rm_{\phi}\gg1$ do not exist. Similarly, 
For small enough $R$ such that $Rm_{\phi}\ll1$, $m_{\phi}$ can be neglected. Going forward for most of this paper we will work in the latter regime neglecting $m_{\phi}$
and also $\phi$ self-interactions.

At the quantum field theory level we have a Dirac fermion $\chi$ with mass $m_{\chi}$ interacting with a scalar $\phi$ through the Lagrange density
\begin{equation}\label{lag}
\mathcal{L} =i\bar \chi \!\not\!\partial \chi - m_\chi \bar \chi \chi - g_\chi \bar \chi \chi \phi +{1 \over 2} \partial \phi \cdot \partial \phi  \ ,
\end{equation}
Without loss of generality we take the Yukawa coupling $g_{\chi}$ to be positive. The Lagrange density in Eq.~(\ref{lag}) has a shift invariance where $g_{\chi} \phi \rightarrow g_{\chi} \phi+c$ and at the same time the parameter $m_{\chi} \rightarrow m_{\chi}-c$. Here $c$ is a constant.

In this paper we  determine the static properties of  bound states with a large number of $\chi$ particles. So we match this  field theory onto a classical theory of $N$  $\chi$ particles interacting with a scalar field $\phi$ by  substituting in the usual particle action the shift invariant mass
\begin{equation}
m_{\chi} \rightarrow m({\bf x}_i)= m_{\chi}+g_{\chi}\phi({\bf x}_i) \ ,
\end{equation}
where ${\bf x}_i (t)$ is the coordinate of the $i$'th $\chi$ particle. 
Then the shift invariant particle Lagrangian becomes
\begin{equation}
\label{lagrangian}
L=-\sum_i m({\bf x}_i)\sqrt{1- {\dot{\bf x}_i^2}}- {1 \over 2}\int d^3x \nabla \phi \nabla \phi  .
\end{equation}
The canonical momenta are,
\begin{equation}
{\bf p}_i =m({\bf x}_i) { {\dot{\bf x}_i} \over \sqrt{1- {\dot{\bf x}_i^2}}}
\end{equation}

The equation of motion for the scalar field is,
\begin{equation}
\label{laplace}
 \nabla^2  \phi({\bf x})=g_{\chi} \sum_i \delta^3({\bf x}-{\bf x}_i){ m({\bf x}_i) \over \sqrt{{\bf p}_i^2+ m({\bf x}_i)^2}} \ .
\end{equation}

For solutions that go to zero at spatial infinity Eq.~(\ref{laplace}) is equivalent to the integral equation,
\begin{equation}
\label{solution}
 \phi({\bf x}) =-g_{\chi} \sum_i {1 \over 4 \pi |{\bf x}-{\bf x}_i|}{m({\bf x}_i) \over \sqrt{{\bf p}_i^2+ m({\bf x}_i)^2}} \ .
\end{equation}
Note that $\phi$ appears on the right hand side of the above differential and integral equations implicitly through the dependence of $m(x)$ on it. 

The Hamiltonian for this system is 
\begin{equation}
\label{energy}
H=\sum_i \sqrt{{\bf p}_i^2+ m({\bf x}_i)^2}-{g_{\chi} \over 2} \sum_i \phi({\bf x}_i){ m({\bf x}_i) \over \sqrt{{\bf p}_i^2+m({\bf x}_i)^2} } \ ,
\end{equation}
where  ${\bf p}_i$ is the momentum of the $i$-th $\chi$ particle and the sum of $i$ is over all the $N$ particles. The second term comes from the scalar part of the field theory Lagrangian. We integrated by parts to put this part of the Lagrangian into a form where the equations of motion can be used. Integrating by parts is not consistent with the $\phi$ shift symmetry since it assumes the field vanishes at spatial infinity. That is why the shift symmetry is not manifest in the second term in Eq.~(\ref{energy}).

Schematically  for a single massless $\chi$ the Hamiltonian in Eq.~(\ref{energy}) has the form,  $H \sim p(1+\sum_i a_i(\phi /p)^i)$. At the quantum field theory level each relativistic $\chi$ particle interaction with the scalar field $\phi$ is suppressed by $1/p$.  This property of scalar interactions with very energetic fermions causes each factor of the scalar field $\phi$, in the classical particle Hamiltonian above to also come suppressed by $1/p$. \footnote[4]{This is very different from N particles of charge $q$ and mass $m$ integrating via an electric field, ${\bf E}=-\nabla \phi^{(em)}$. Then the static Hamiltonian is
\begin{equation}
H=\sum_i \sqrt{{\bf p}_i^2+ m^2}+{q \over 2} \sum_i \phi^{(em)}({\bf x}_i) \ .
\end{equation}}

Since we are interested in a large number of fermions, $N>>1$, throughout this paper we replace  sums over particles by an integral over a  phase space density $f({\bf r},{\bf p})$ of a Fermi gas, {\it i.e.}, 
\begin{equation}
\sum_i  \rightarrow  \int {d^3r} \int {d^3p \over (2 \pi)^3}f({\bf r},{\bf p}). \nonumber
\end{equation} 
We also assume spherical symmetry. This corresponds to having the  $\chi$ particles confined to a coordinate sphere of radius $R$, and at each spatial point having a Fermi sea filled to  a Fermi momentum $p_F(r)$ that can depend on the radial coordinate $r$, 
\begin{equation}
f({\bf r},{\bf p})=2\theta (R-r) \theta (p_F(r)-p) \ , 
\end{equation}
where the factor of 2 is from the spin degrees of freedom. 

In this case, the total number of particles
\begin{equation}
\label{number}
N=\sum_i 1 =  {4 \over 3\pi}\int_0^R r^2 dr \left(p_F(r)\right)^3 \ .
\end{equation}
Converting the sums over $i$ in Eq.~(\ref{energy})  to phase space integrations we find that the total energy of such a Fermi gas with $N$ $\chi$ particles and radius $R$ is

\begin{equation}\label{etot}
E\left(N,R\right)={4 \over \pi} \int_0^R r^2 dr \left[m(r)^4 h\left(p_F(r)/|m(r)| \rule{0mm}{3.6mm} \right)
-{1 \over 2} g \phi (r) m(r)^3  i\left(p_F(r)/|m(r)|\rule{0mm}{3.6mm}\right) \right] \ ,
\end{equation}
where the integrations over momenta gave rise to the functions
\begin{eqnarray}
h(z)&=&\int_0^z du {u^2\sqrt{1+u^2}}={1 \over 4} \left( i(z) +z^3 \sqrt{1+z^2} \right) \ , \\
i(z)&=&\int_0^z dz {u^2 \over\sqrt{1+u^2}}={1 \over 2}z\sqrt{1+z^2} -{1 \over 2}{\rm arcsinh}(z) \ .
\end{eqnarray}
For small $z$,  $i(z) \sim z^3/3$  while for large $z$, $i(z) \sim z^2/2$.

Inside the nugget the scalar field $\phi(r)$ satisfies the differential equation,
\begin{equation}
\nabla^2\phi(r) = \left(\frac{d^2}{dr^2} + \frac{2}{r} \frac{d}{dr} \right)\phi(r)= \frac{g_\chi}{\pi^2} m(r)^3  i \left(p_F(r)/|m(r)| \rule{0mm}{3.6mm}\right) \ . \label{lap}
\end{equation}
At the origin the scalar field satisfies the boundary condition $\phi'(0)=0$. Outside the nugget
\begin{equation}
\phi(r)=\phi(R){R\over r} \ .
\end{equation}
Differentiating this gives the boundary condition, $\phi'(R)=-\phi(R)/R$, at $r=R$.

With $m_{\phi}$ neglected the integral over the angle between the ${\bf r}$ and ${\bf r'}$ (the position of the $\chi$ particles that source the field $\phi({\bf r})$) can be done in Eq.~(\ref{solution}) leaving just the radial integral,
\begin{equation}
\phi(r) = - \frac{2 g_\chi}{\pi} \left[ \frac{1}{r} \int_0^r dr' \, r^{\prime 2} m(r')^3 i \left(p_F(r')/|m(r')| \rule{0mm}{3.6mm}\right) + \int_r^R dr'\, r'\, m(r')^3 i \left(p_F(r')/|m(r')| \rule{0mm}{3.6mm}\right) \right] \ .
\end{equation}
Differentiating the above equation with respect to $r$ gives for $r<R$,
\begin{equation}\label{proof}
\frac{d m(r)}{d r} = \frac{2 g^2_\chi}{\pi r^2} \int_0^r dr' \, r^{\prime 2} m(r')^3 i \left(p_F(r')/|m(r')| \rule{0mm}{3.6mm}\right) \ .
\end{equation}
As was noted earlier the effective mass, $m(r) \equiv m_\chi + g_\chi \phi(r)$.  The field $\phi(r)$ is negative and gets larger in magnitude as one goes towards the center of the nugget at $r=0$. Suppose at the center $m(0)$ is positive then the above differential equation indicates that $m(r)$ increases as one increases  $r$ towards the surface of the nugget. Conversely if $m(0)$ is negative then it decreases as $r$ increases towards the surface of the nugget. {\em Hence the effective mass never changes sign inside the nugget.} In practice for the situations we consider this means that  $m(r)$ is always positive.

The properties of a nugget can be calculated once the $r$-dependence of the Fermi momentum $p_F(r)$ is known. In general, it can be determined from the hydrostatic equilibrium. We describe this method in Sec.~\ref{HE} but find it difficult to implement numerically. So in this paper we take a more heuristic approach making an ansatz for the form of $p_F(r)$ and then use Eq. ({\ref{number}) to write $p_F(r)$ as a function of  $r$, $N$ and $R$.  For each $N$ and $R$ the non-linear differential equation for $\phi(r)$ ({\it i.e.}, Eq.~(\ref{lap})) is solved using the boundary conditions on the radial derivative of $\phi$ at the origin and the surface of the nugget. This is used to compute $E(N,R)$ which is then minimized with respect to $R$ at fixed $N$ to find the nugget size and binding energy. In the non-relativistic regime where the equations of hydrostatic equilibrium can be solved this heuristic approach gives reasonable results for the two ansatzs for the Fermi momentum that we will make.

A very simple ansatz the Fermi momentum inside the nugget is a constant independent of $r$
\begin{equation}\label{ansatz0}
p_F(r)=\left(9 \pi N \over 4\right)^{1/3}{1\over R} \ .
\end{equation}
 Constant Fermi momentum in the nugget is not compatible with the physical condition that the Fermi momentum be continuous at the surface of the nugget. 
A one-parameter family of physically more reasonable ansatzs with this property  is 
\begin{equation} 
\label{ansatza}
p_F(r)=\left({3 \pi\Gamma(4+3a) N \over 8\Gamma(1+3a)}\right)^{1/3}  \left(1-{r\over R}\right)^{a}{1 \over R} \ .
\end{equation}
This form of $p_F(r)$ with $a=1/2$ is a good approximation for the solution to the Lane-Emden equation for hydrostatic equilibrium which describes the bound states in non-relativistic case~\cite{Wise:2014jva}.
Motivated by this, we will use $a=1/2$ for both non-relativistic and relativistic regimes.

\section{Hydrostatic Equilibrium}\label{HE}

In this section, we briefly discuss the approach to the Yukawa bound state problem based on hydrostatic equilibrium. As discussed in Sec.~\ref{method}, the key is the knowledge of the Fermi momentum as a function of the position, $p_F(r)$.
It can be determined by the the energy-momentum conservation law $\partial_{\mu}T^{\mu\nu}=0$, where $T^{\mu\nu}$ is the stress-energy tensor.

In the static situation we are dealing with, the above equation becomes $\nabla^kT^{kl}=0$, and the spatial components are
\begin{equation}
T^{kl}({\bf r})=\sum_{i}{ p_i^k p_i^l \over \sqrt{ {\bf p}_i^2+m({\bf x}_i)^2}}\delta^3({\bf r}-{\bf x}_i)-\nabla^k\phi({\bf r})\nabla^l \phi({\bf r})+{\delta^{kl} \over 2} \nabla\phi({\bf r}) \cdot \nabla \phi({\bf r}) \ .
\end{equation}
Conservation of stress-energy, $\nabla^kT^{kl}=0$, then implies the first order integral-differential equation,
\begin{equation}
\left({r^2 \over 3 \pi^2}\right)p_F^{\prime} (r) {p_F(r)^4 \over \sqrt{m(r)^2+p_F(r)^2}}= - {g_{\chi} \over \pi^2}\nabla^2\phi(r)\int_0^r dr' r'^2 m(r')^3i(p_F(r')/m(r')) \ ,
\end{equation}
where $p_F^{\prime}(r)= dp_F(r)/dr$. Since $\nabla^2\phi(r)$ is positive we see that $p_F^{\prime}(r)$ is negative. {\it Hence the Fermi momentum increases as $r$ decreases}.

Converting this integral-differential equation into a second order differential equation one obtains,
\begin{equation}
\label{hydro}
\frac{1}{r^2}\frac{d}{dr} \left[ \frac{1}{\left( \nabla^2\phi \right) } \frac{r^2}{3\pi^2} \frac{( p_F^{\prime}(r)) p_F(r)^4}{\sqrt{p_F(r)^2+ m(r)^2}} \right]= - \frac{g_\chi}{\pi^2} m(r)^3 i(p_F(r)/m(r)) \ ,
\end{equation}
as the equation of hydrostatic equilibrium. This is a second order differential equation  and the  boundary conditions are $p_F'(0)=0$ and $p_F(R)=0$. 

In general, the equation of hydrostatic equilibrium Eq.~(\ref{hydro}) cannot be solved independently --- it is coupled to the Laplacian equation Eq.~(\ref{lap}), and the two differential equations must be solved together in order to determine $p_F(r)$ and $\phi(r)$. For each radius $R$, $p_F(r)$ fixes the $\chi$ number $N$ through Eq.~(\ref{number}), and determines the total energy $E$ through Eq.~(\ref{etot}).

In the non-relativistic limit $p_F(r) \ll m(r)$, the above equation can be simplified. The number density of $\chi$ particles is, $n(r)=p_F(r)^3/(3 \pi^2)$ and in this limit the Laplacian of $\phi$ is,
$\nabla^2 \phi(r)=g_{\chi} n(r)$. Introducing the effective pressure for non-relativistic Fermi gas, $p(r)=p_F(r)^5/(15 \pi^2 m(r))$, the equation for hydrostatic equilibrium takes the more recognizable form,
\begin{equation}
\frac{1}{r^2}\frac{d}{dr}\left[ {1 \over n(r)} r^2 p'(r) \right]=-g_{\chi}^2 n(r) \ .
\end{equation}

Under the further assumption of weak field, $g\phi\ll m_\chi$, the solution to the equation of hydrostatic equilibrium is known and this solution was discussed in~\cite{book}. The application to dark matter bound states was discussed in~\cite{Wise:2014jva}. We find that in the non-relativistic weak field regime, the choice $a=1/2$ in Eq.~(\ref{ansatza}) is  a very good approximation and as was remarked on in~\cite{Wise:2014jva} even the constant Fermi-momentum ansatz in Eq.~(\ref{ansatz0}) gives results that have the correct qualitative behavior.

\section{Perturbation Theory}

The main purpose of this section is to show how perturbation theory in the coupling breaks down for large $N$. The problem with perturbation theory does not depend on the explicit ansatz for the $r$ dependence of the Fermi momentum and so in this section we use the very simple constant Fermi momentum ansatz, $p_F(r)=p_F \theta(R-r)$. Expanding in powers of $g_{\chi}$ 
\begin{equation}
\phi({\bf x})=\phi_0({\bf x})+\phi_1({\bf x})+\ldots \ ,
\end{equation}
where the term with subscript $n$  is of order  $g_{\chi}^{2n+1}$.
Neglecting $m_{\phi}$ we have from eq.(\ref{solution}) that,
\begin{equation}
\phi_0({\bf x})=-g_{\chi} \sum_i {1 \over 4 \pi |{\bf x}-{\bf x}_i|}  {m_{\chi}  \over \sqrt{{\bf p}_i^2+m_{\chi}^2}} \ .
\end{equation}
Using our ansatz for $p_F(r)$ this becomes,
\begin{equation}\label{phi0}
g_\chi \phi_0(r) = -\alpha_\chi {3 N \over 2 R}\left(3 - {r^2 \over R^2}\right) \left({m_{\chi} \over p_F} \right)^3 i(p_F/m_{\chi}) \ ,
\end{equation}
where the function $i(z)$ was defined in the previous section.

\begin{figure}[t]
\centerline{\includegraphics[width=0.9\columnwidth]{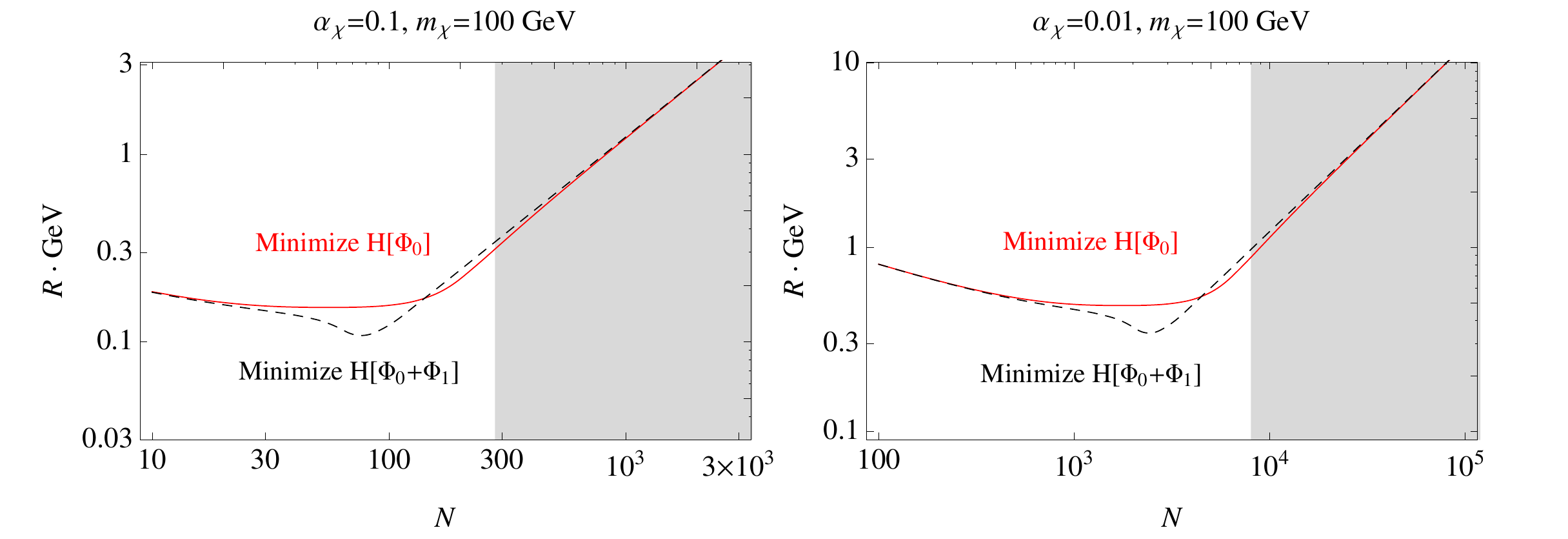}}
%\centerline{\includegraphics[width=1\columnwidth]{ratios.pdf}}
\caption{Radius of the nugget from analytic perturbative solution. The solid and dashed curves are solutions obtained by minimizing $E(N,R)$ with $\phi_0$ and $\phi_0+\phi_1$ (from Eqs.~(\ref{phi0}) and (\ref{phi1})), respectively. In the shaded region, the effective mass $m(r)$ becomes negative.}\label{f1}
\end{figure}

\begin{figure}[t]
\centerline{\includegraphics[width=0.9\columnwidth]{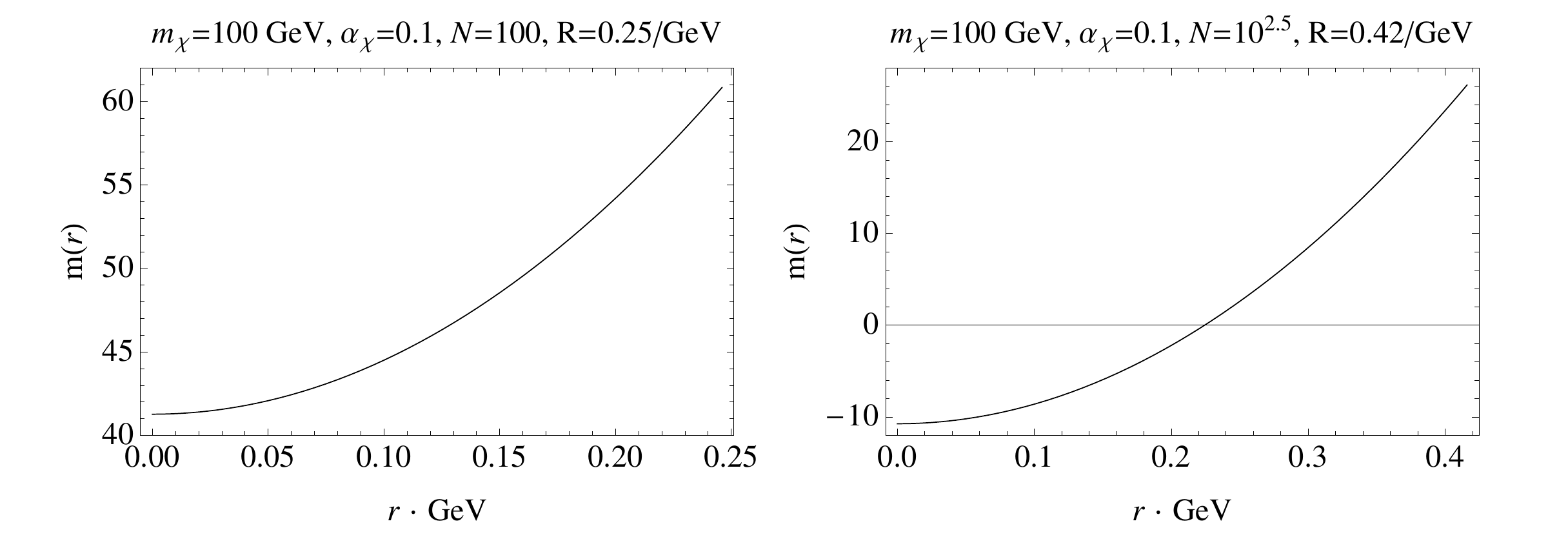}}
%\centerline{\includegraphics[width=1\columnwidth]{ratios.pdf}}
\caption{The effective mass $m(r)=m_\chi+g\phi(r)$ as a function of position $r$ inside the nugget, for two sets of parameters. The radius $R$ is chosen to minimize the total energy $E(N,R)$. In the left plot, $m(r)$ is always positive, while in the right plot $m(r)$ changes sign.}\label{f3}
\end{figure}

At the next order,
\begin{equation}
\phi_1({\bf x})=-g_{\chi}^2 \sum_i {1 \over 4 \pi |{\bf x}-{\bf x}_i|}{\phi_0 ({\bf x}_i){\bf p}_i^2 \over ({\bf p}_i^2+m_{\chi}^2)^{3/2}} \ ,
\end{equation}
which with our ansatz for $p_F(r)$ becomes,
\begin{equation}\label{phi1}
g_\chi \phi_1(r) =\alpha_\chi^2 {27N^2 \over 40 m_\chi R^2}\left(5- {r^2 \over R^2}\right)^2 \left({m_{\chi} \over p_F} \right)^6 i(p_F/m_{\chi}) j(p_F/m_{\chi}) \ ,
\end{equation}
where the function
\begin{equation}
j(z)=\int_0^z d u{ u^4 \over (1+u^2)^{3/2}}={z (3+z^2) \over 2 \sqrt{1+z^2}}-{3 \over 2}{\rm arcsinh}(z) \ .
\end{equation}

With the perturbative solution for $\phi$ the total energy $E(N,R)$ can be obtained using Eq.~(\ref{etot}). Minimizing it with respect to $R$ (at fixed $N$) yields the radius of the ground state of $N$ $\chi$ particles and the binding energy of that state. We find as one increases $N$ the radius first shrinks and then expands as shown in Fig.~\ref{f1} where we used  $m_{\chi}=100{\rm GeV}$ and $\alpha_{\chi}$ equal to $0.1$ and $0.01$.

The hierarchy  $ \phi_0 \gg  \phi_1$ is satisfied throughout the nugget. Despite this, perturbation theory breaks down for large $N$ because the field becomes very large towards the core of the nugget driving $m(r)$ from positive to negative values (the dashed curves in Fig.~\ref{f3}) which is in conflict with the general results we derived earlier. In other words, as $\phi_0$ gets more negative, there is a region of $r$ where $m_\chi + g\phi_0(r)$ is very small and even though $\phi_1\ll \phi_0$, $m_\chi + g \phi_0(r)+g \phi_1(r)$ is not close to $m_\chi + g\phi_0(r)$.

\section{Numerical Approach}\label{numerical}

Since solving the hydrostatic equilibrium equations is too difficult, and perturbation theory is not valid for large $N$ because the field $\phi$ gets too large, we adopt the method described in Sec.~\ref{method}. For fixed $N$, the Laplacian equation (\ref{lap}) is solved numerically, for different choices of $R$. We present results for both the constant and power law (with $a=1/2$) ansatzs for $p_F(r)$ given in Eqs.~(\ref{ansatz0}) and (\ref{ansatza}) respectively. The energy in Eq.~(\ref{etot}) is minimized as a function of $R$ to determine the radius and binding energy of a nugget containing $N$ $\chi$ particles. 

\begin{figure}[t!]
\centerline{\includegraphics[width=1.0\columnwidth]{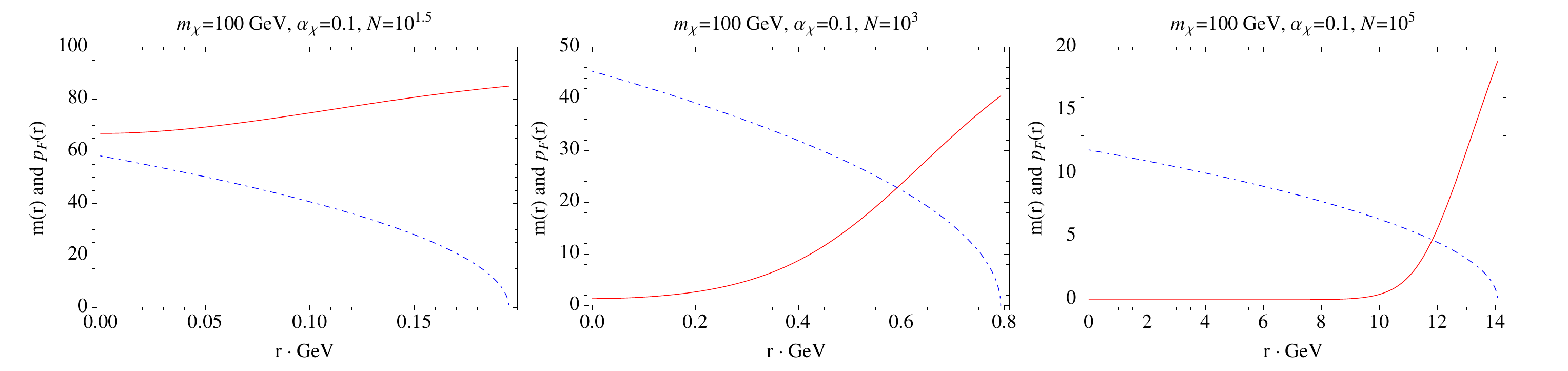}}
\centerline{\includegraphics[width=1.0\columnwidth]{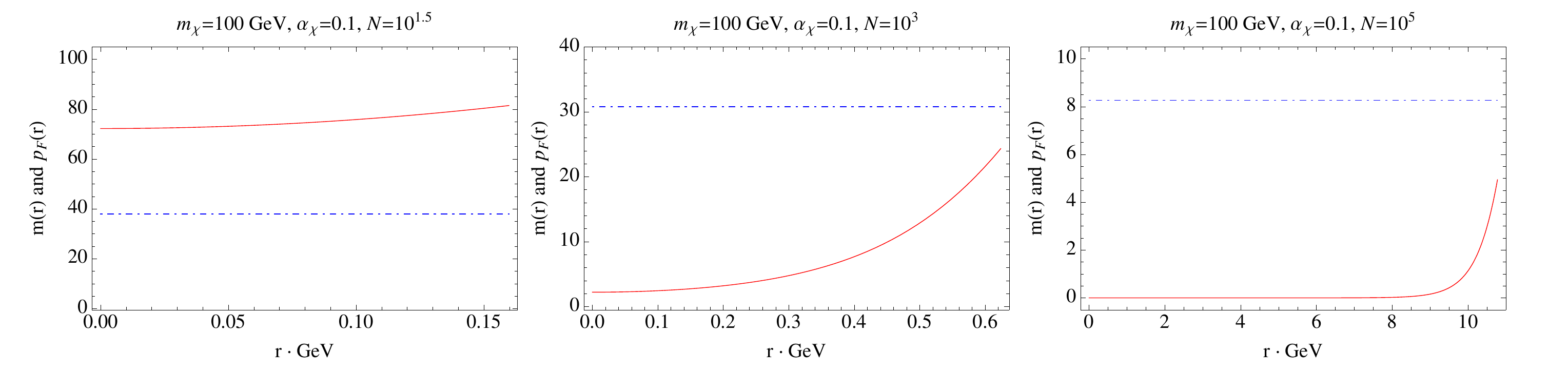}}
\caption{Results of the numerical approach described in Sec.~\ref{numerical}: effective mass $m(r)$ and the fermi momentum $p_F$ as a function of $r$ inside the nugget.
The parameters used are $\alpha_\chi=0.1$, $m_\chi=100\,$GeV. The radius $R$ is chosen to minimize the total energy $E(N,R)$. 
We used the two different ansatzes for $p_F(r)$  in Eq.~(\ref{ansatza}) (first row) and (\ref{ansatz0}) (second row).
}\label{f4}
\vspace{0.5cm}
%\end{figure}
%\begin{figure}[t]
\centerline{\includegraphics[width=1.0\columnwidth]{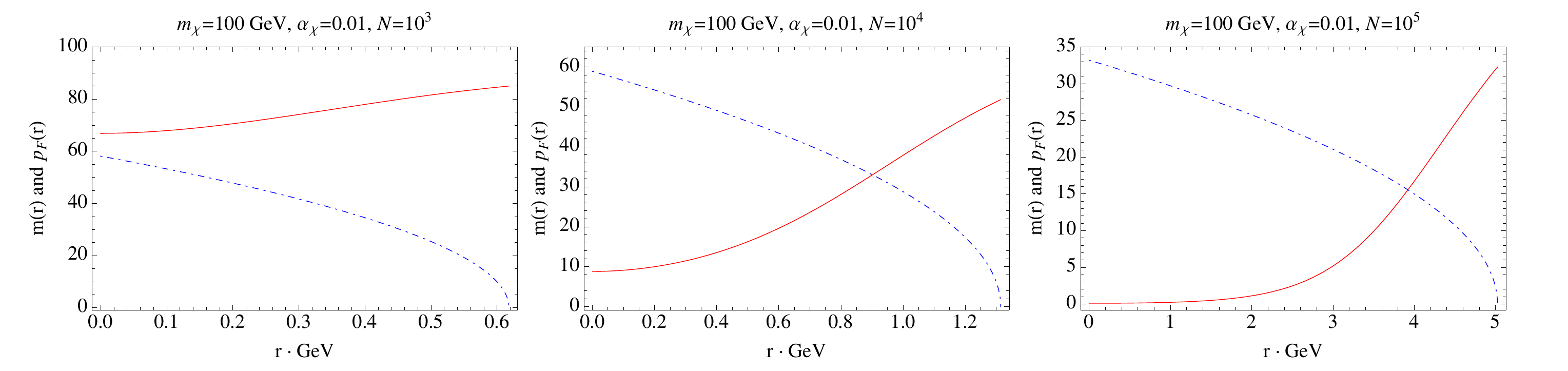}}
\centerline{\includegraphics[width=1.0\columnwidth]{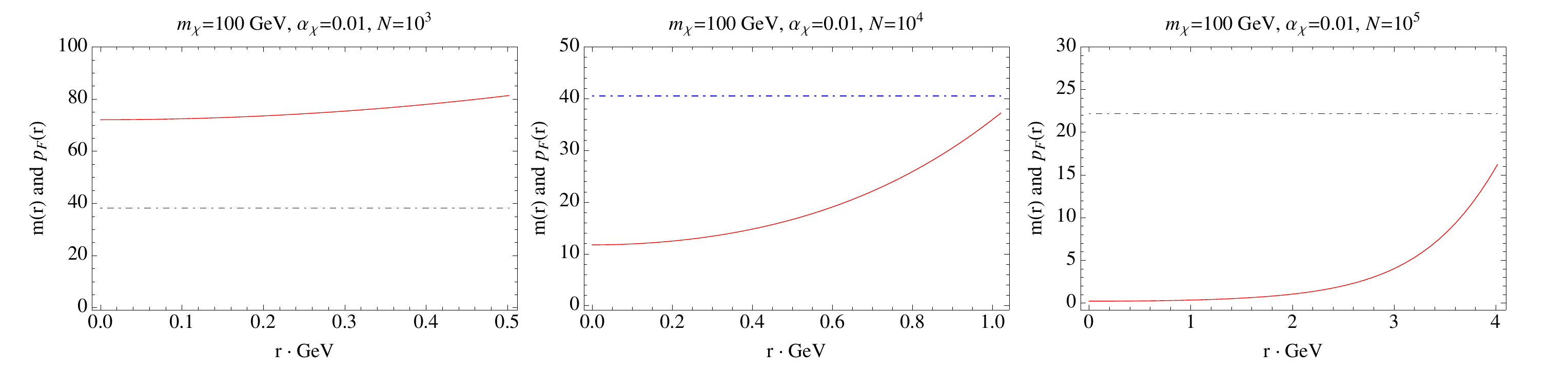}}
\caption{Same as Fig.~\ref{f4}, but with $\alpha_\chi=0.01$.}\label{f4.5}
\end{figure}

The model parameters that determine the physics of nuggets are, $m_{\chi}$, $\alpha_{\chi}=g_{\chi}^2/(4 \pi)$ and the mediator mass $m_{\phi}$. Our ansatzs for the dependence of the Fermi momentum on the radius do not introduce any new dimensionful parameters once they are normalized to the number of particles. We are neglecting $m_{\phi}$ here and so the only dimensionful parameter is $m_{\chi}$. Hence we work at the fixed value, $m_{\chi}=100{\rm GeV}$. Using dimensional analysis we can determine the dependence of physical quantities on $m_{\chi}$, for example  $R \propto 1/m_{\chi}$ .  To capture the trends with $\alpha_{\chi}$ we display our results for two values $0.1$ and $0.01$. 

\begin{figure}[t]
\centerline{\includegraphics[width=0.425\columnwidth]{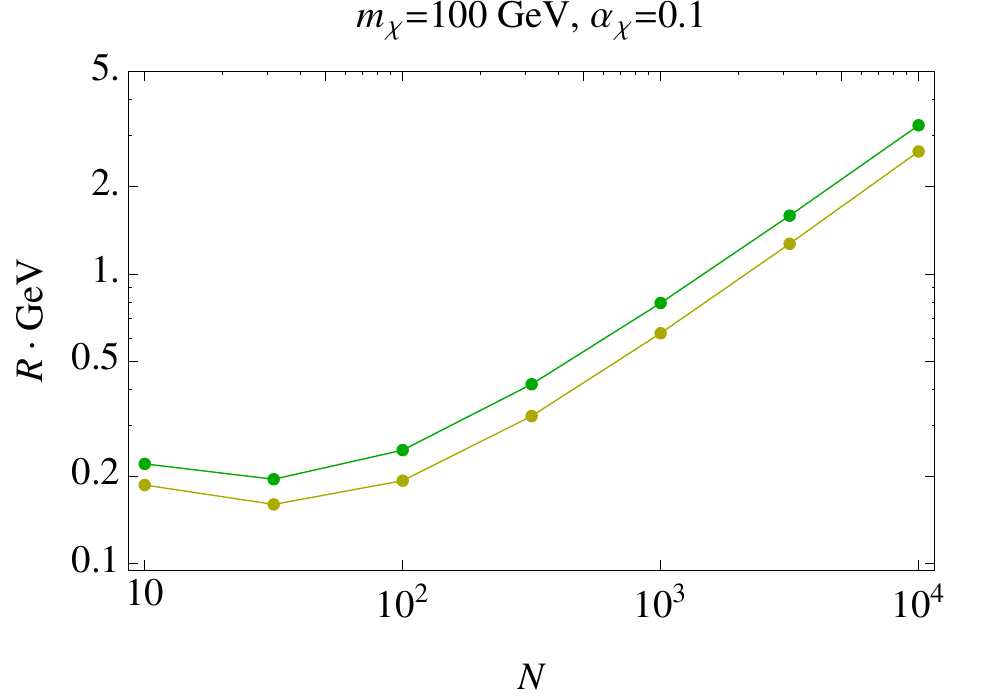}
\includegraphics[width=0.425\columnwidth]{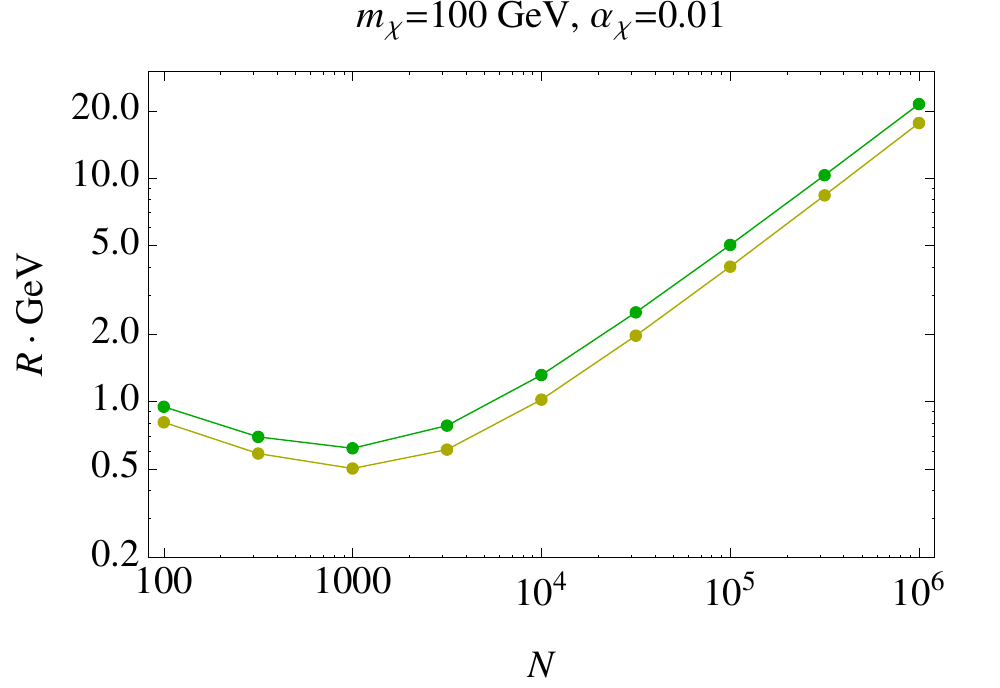}}
\vspace{0.3cm}
\centerline{\includegraphics[width=0.425\columnwidth]{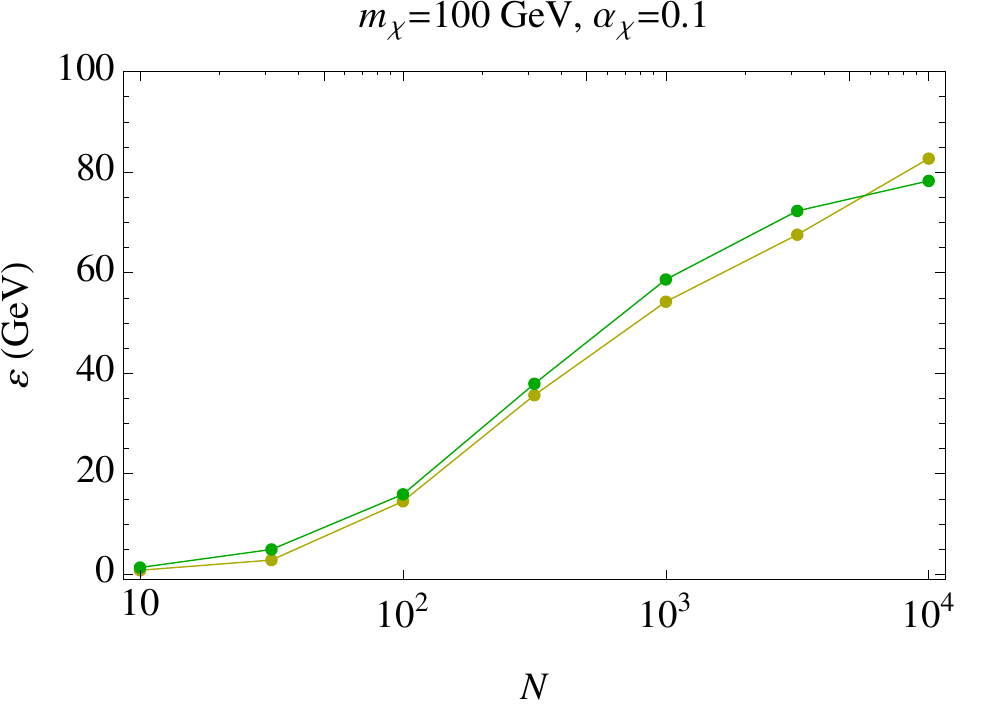}
\includegraphics[width=0.425\columnwidth]{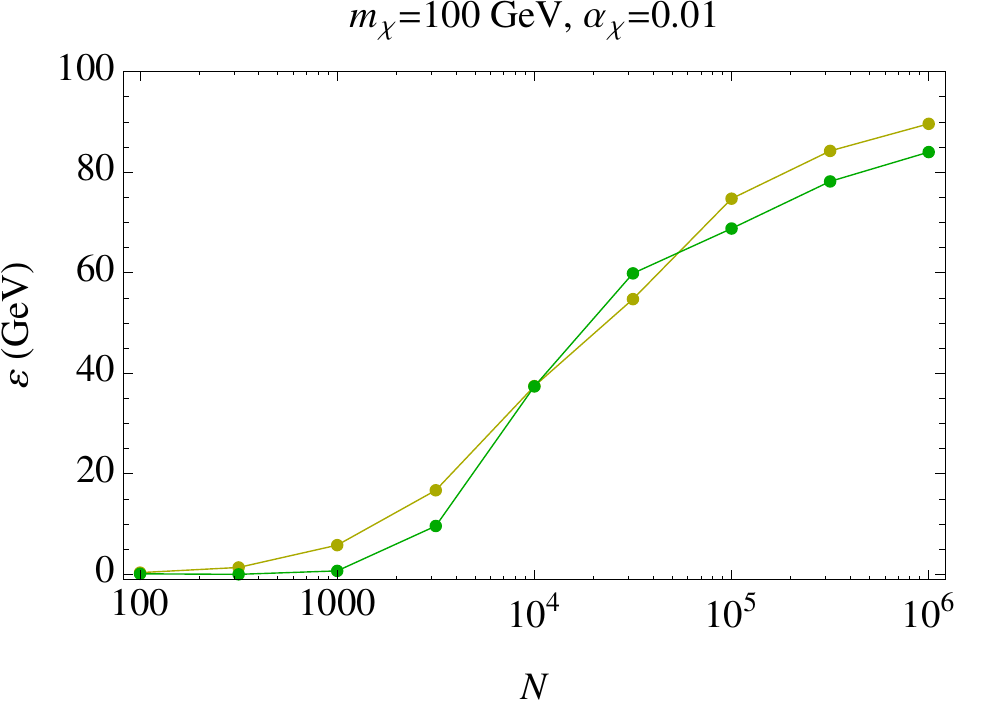}}
%\centerline{\includegraphics[width=1\columnwidth]{ratios.pdf}}
\caption{Properties of the nuggets containing $N$ $\chi$ particles using the numerical approach described in Sec.~\ref{numerical}. The nugget radius $R$ versus $N$ (first row) and binding energy per particle $\varepsilon$ versus $N$ (second row) are solved using the numerical method discussed in Sec.~\ref{numerical}. The green (darker) and yellow (lighter) curves correspond respectively to the ansatzs for $p_F(r)$ in Eqs.~(\ref{ansatz0}) and (\ref{ansatza}).
}\label{f2}
\end{figure}

In Fig.~\ref{f4} and \ref{f4.5}, we show $m(r)$ and $p_F(r)$ throughout the nugget, for different values of $N$. We find that for small $N$ the $\chi$ particles are non-relativistic throughout the bound state. For larger $N$, the $\chi$ particles are not ultra relativistic near the surface, but are ultra relativistic in the core region. The effective mass $m(r)$ becomes small near the core but does not change sign.

In Fig.~\ref{f2}, we plot the nugget radius $R$ and binding energy per particle,
\begin{equation}
\varepsilon\equiv m_\chi - E(N,R)/N \ , 
\end{equation}
as a function of the number of particles $N$, for the same sets of parameters as Fig.~\ref{f4}.
Note that as $N$ increases, the radius $R$ first shrinks and then grows. 
At larger $N$, more of the $\chi$ particles are relativistic, the Yukawa interactions among these particles are $m/p$ suppressed, and the Fermi pressure pushes the minimal energy configuration to larger $R$. The binding energy per particle $\varepsilon$ increases monotonically with $N$, indicating that the nuggets are stable against breaking up into smaller pieces.
For very large $N$, the binding energy per particle $\varepsilon$ is expected to reach a plateau because $\varepsilon$ cannot exceed $m_\chi$.
The behavior of $R$ and $\varepsilon$ as functions of $N$ are the main results of this paper.

The qualitative behavior of our results for the $N$ dependence of the nugget radius and binding energy do not depend on the the particular ansatz for the Fermi momentum chosen. The most striking feature, that the radius first decreases with $N$ and then increases, occurs for both ansatzs for $p_F(r)$ that we used and is even present if perturbation in the coupling $\alpha_{\chi}$ is used to determine $\phi(r)$.

\begin{figure}[t!]
\centerline{\includegraphics[width=0.95\columnwidth]{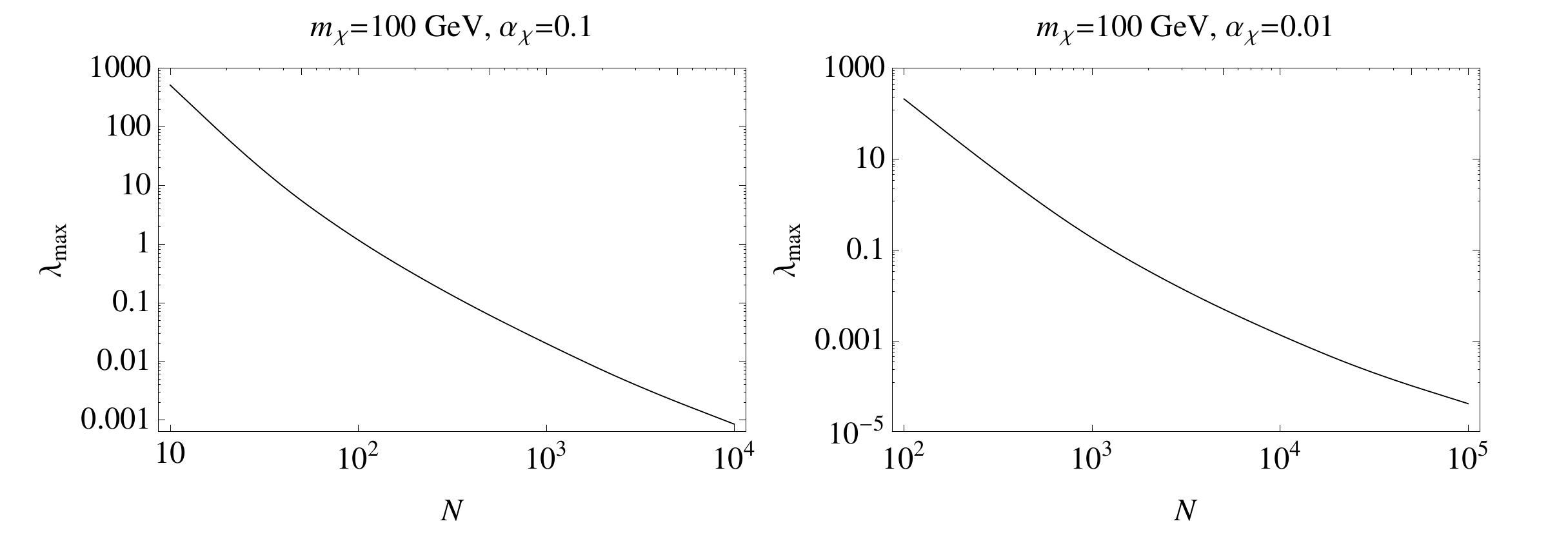}}
%\centerline{\includegraphics[width=1\columnwidth]{ratios.pdf}}
\caption{Upper bound on $\lambda$ as defined in Eq.~(\ref{max}) from numerical approach.
}\label{f6}
\end{figure}

Throughout this paper, we have neglected self-interactions of the $\phi$ field. We can estimate the range of self-couplings $\lambda$ for which this is a reasonable approximation. Suppose the Hamiltonian in Eq.~(\ref{energy}) contains an additional term $\delta H[\phi] =  \int d^3 x \lambda \phi(x)^4/4 ! $.
Using our solution for $\phi$, we show in Fig.~\ref{f6} (for $\alpha_{\chi}=0.1$ and $0.01$) the largest values of $\lambda$ for which this new contribution does not to exceed the total binding energy we calculated before,
\begin{equation}\label{max}
\lambda_{\rm max} (N) = \left|\frac{N m_\chi - E(N,R)}{\delta H[\phi]} \right|  \ .
\end{equation}
Here  the $a=1/2$ power law ansatz for $p_F(r)$ in Eq.~(\ref{ansatza}) was used. Clearly for larger $N$ this requires smaller $\lambda$ for the self interaction contribution to be negligible.
A non-zero $\lambda$ term draws the $\phi$ field closer to 0, thus it tends to reduce the nugget radius $R$.

To close this section, we discuss the possible existence of nuggets with radius $R$ much larger than the screening length $1/m_\phi$. 
Even for very large $R$ there is a surface attraction that may cause the nugget to continue to grow as $N$ increases.
Idealizing the surface of a large nugget as flat and containing constant number density $n$ of $\chi$ particles, the potential that attracts a $\chi$ particle at hight $z$ above the surface is
\begin{equation}
V(z) = - \frac{2 \pi \alpha_\chi n}{m^2} e^{-m_\phi z} \ .
\end{equation}

%It seems intuitively clear that the screening forbids such states but it is not difficult to address this issue somewhat quantitatively using the methods developed in this paper.  Assuming the constant Fermi momentum ansatz  $p_F \sim N^{1/3}/R$, that the $\chi$ particles are non-relativistic and that the weak field limit is appropriate,  the sum of the total kinetic and potential energy for a nugget with $R\gg1/m_{\phi}$ is
%\begin{equation}
%\label{largenugget}
%E(N,R)-Nm_{\chi} = a {N^{5/3} \over m_{\chi} R^2} - b{ \alpha_{\chi}N^2 \over R^3 m_{\phi}^2} \ .
%\end{equation}
%Here $a$ and $b$ are positive constants. The first term in Eq.~(\ref{largenugget}) arises from the kinetic energy and the second from the potential energy.  For fixed $N$, the above expression is minimized at $R=0$ or $\infty$ and hence stable nuggets with $R\gg1/m_{\phi}$ do not exist. 

\section{Application to Dark Matter}

\begin{figure}[b!]
\centerline{\includegraphics[width=0.9\columnwidth]{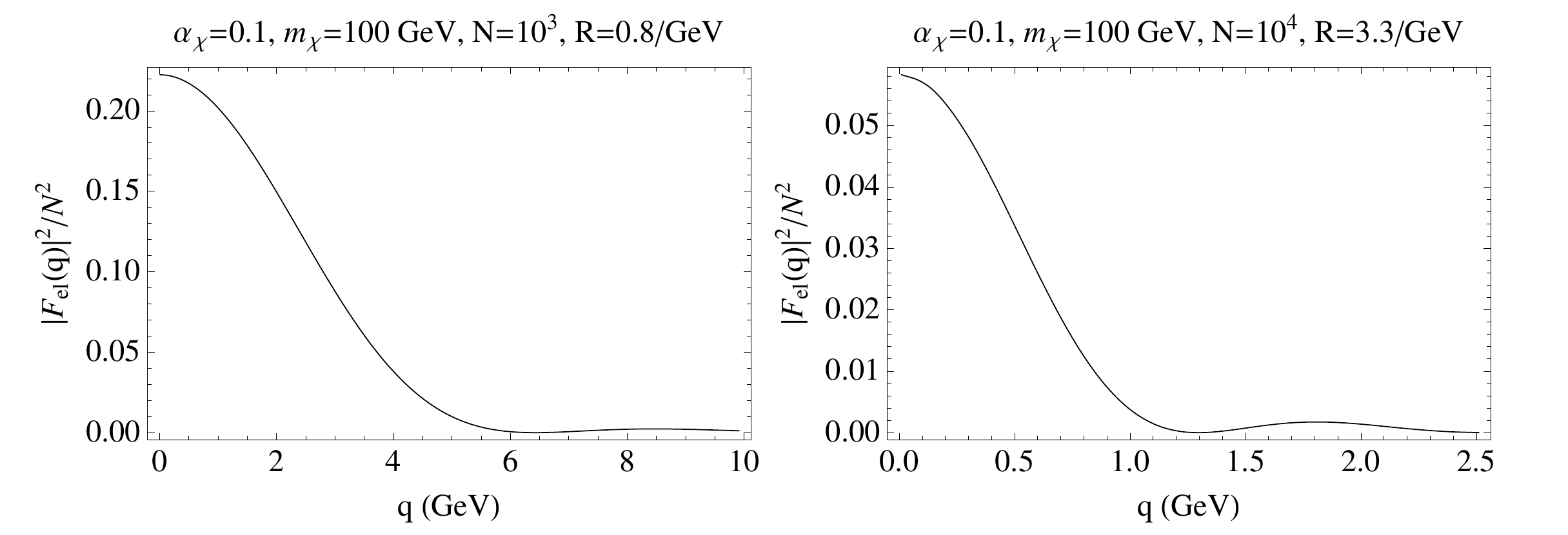}}
\centerline{\includegraphics[width=0.9\columnwidth]{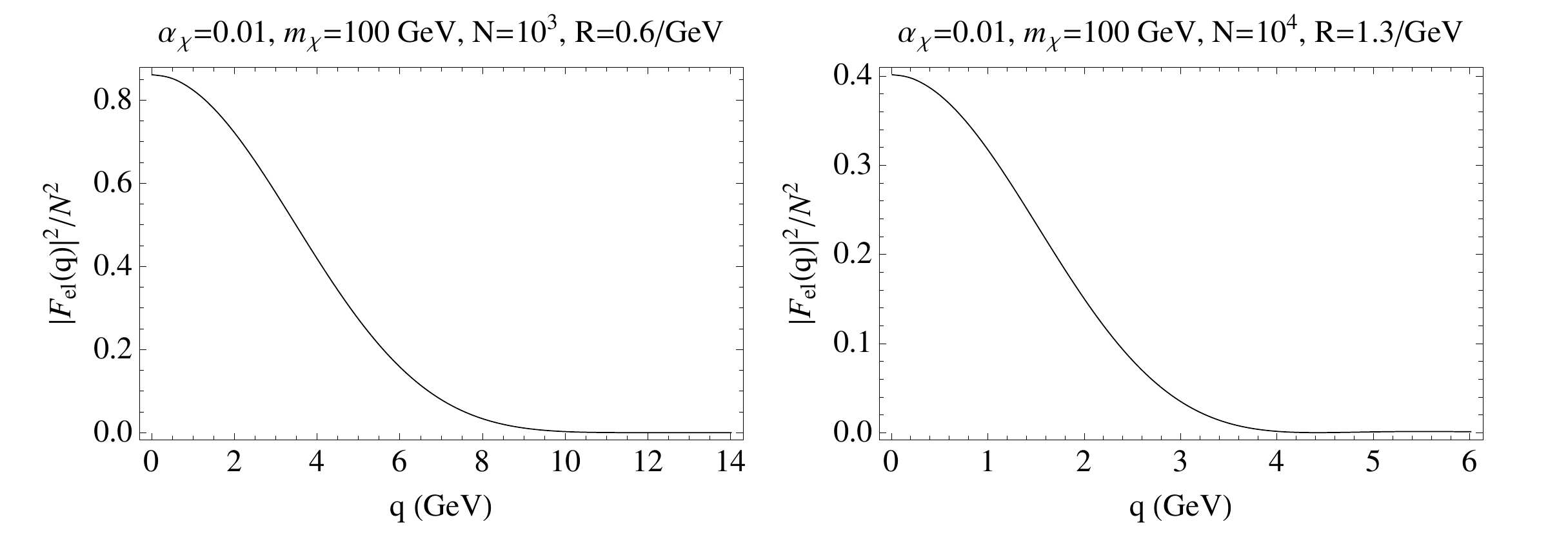}}
\caption{Ratio of the elastic form factor in nugget scattering to $N^2$, as a function of the momentum transfer $q$. Here we have used the $a=1/2$ power law ansatz on $p_F(r)$ in Eq.~(\ref{ansatza}).}\label{f5}
\end{figure}

If the $\chi$ particles are the DM, then for a range of parameters nuggets can form in the early universe~\cite{Wise:2014jva}, shortly after the (free) $\chi$'s freeze out. Such dark nuggets can be cosmologically abundant if the DM relic density is dominated by $\chi$ particles, {\it i.e.}, the DM is asymmetric. 
In the example shown in Fig.~\ref{f2}, it is possible to have a bound state containing thousands of weak scale $\chi$ particles --- a nugget with mass of order $\sim 10^3$\,TeV.
Thermal supermassive DM cannot have the correct relic density because the unitarity bound indicates that the annihilation cross section is too small. 
However, our analysis of Yukawa bound states provides an interesting way to have DM that is effectively supermassive.

The rate for their direct detection is determined partly by the elastic form factor $F(q^2)$,
\begin{equation}
F(q^2) = \sum_{i=1}^N e^{i {\bf q} \cdot {\bf x}_i} { m({\bf x}_i) \over \sqrt{{\bf p}_i^2+m({\bf x}_i)^2} }= \frac{4}{\pi} \int_0^R r^2 dr \frac{\sin (qr)}{qr}  (m(r))^3 i(p_F(r)/m(r)) \ .
\end{equation}
For $q=0$, the naive expectation for the form factor is  $F(0)=N$. However relativistic particle couplings to the scalar mediator are suppressed and this causes  $F(0)$ to be  less than $N$. This feature becomes more prominent for large $N$ since then more of the $\chi$ particles are relativistic. The form factor falls as the momentum transfer $q^2$ increases because the scattering becomes less coherent. These features are shown in Fig.~\ref{f5}. This form factor will also be relevant for elastic DM-DM scattering through virtual $\phi$ particle exchange.

In this paper we have focussed on the lowest energy bound state of $N$ $\chi$ particles. Of course there are excitations and so inelastic processes are possible. We have not explored the form factors relevant in that case but expect that  the inelastic channels are suppressed at small $q$ because of Pauli blocking.

Indirect detection signals of DM bound states are also very important. Since the binding energy per $\chi$ particle grows with the number $N$, the $\chi$'s are more deeply bounded in larger nuggets. There will be a release of energy by emission of a mediator $\phi$ (either real or virtual) when a free $\chi$ particle is captured by a nugget or when a small nugget captured by a large one. In the model discussed in Ref.~\cite{Wise:2014jva}, a real mediator $\phi$ materializes as a SM $\mu^+\mu^-$ or $\pi\pi$ final state. This could offer interesting signals for indirect DM searches using cosmic rays.

\section{Outlook} 
 
We have studied  bound states of  a large number of Dirac fermions $\chi$ interacting through exchange of a light scalar field that they are Yukawa coupled to. For very large $N$ the  cores of these objects contain very relativistic $\chi$'s and the size of the state $R$ increases with $N$. That is in contrast to the smaller $N$ regime where the $\chi$ particles are non relativistic and the size of the state shrinks as $N$ increases. There are several extensions of this work that are worth pursuing. 

We made a number of approximations in order to draw these conclusions that are worthy of further exploration. For example we used a simple Fermi gas model for the fermions. For strong enough coupling pairing of fermions and Bose-Einstein condensation may occur. Also we neglected the scalar self coupling. A preliminary estimate we made shows that in some cases the scalar self coupling must be very small for this to be  a good approximation and so it would be nice to extend our analysis to include the scalar self coupling. 
%We have shown that nuggets with $Rm_\phi\gg1$  are not stable. 
It would be interesting to explore  in more detail the properties of nuggets in the regime $Rm_\phi\gtrsim1$.

Finally for the interpretation of $\chi$ particles as dark matter it is important to estimate the fraction of dark matter that ends up in bound states with $N>2$,\footnote[4]{The cosmological production of two body bound states was considered in~\cite{Wise:2014jva}.} {\it i.e.}, the analog of big-bang nucleosynthesis in the dark sector.
For recent progress along this direction in other models, see~\cite{Krnjaic:2014xza} and \cite{Detmold:2014qqa}.

\section*{Acknowledgement}
This work started from discussions with Natalia Toro who pointed out the suppression at high energies of Yukawa couplings and the form of the effective classical particle theory in Eq.~(\ref{lagrangian}). We are grateful to her for these key insights.
We also thank Peter Graham and Jeremy Mardon for useful discussions.
This work is supported by the Gordon and Betty Moore Foundation through Grant
No.~776 to the Caltech Moore Center for Theoretical Cosmology and Physics, and by the DOE Grant DE-SC0011632, and also by a DOE Early Career Award under Grant No. DE-SC0010255. We are also grateful for the support provided by the Walter Burke Institute for Theoretical Physics.

\end{document}